\documentclass{article}

\usepackage{psfig}

\begin{document}

\sloppy
\baselineskip=18pt

\begin{verbatim}
                                         Poster paper presented at JENAM-2000
                                         Section 2: "Morphology and dynamics of stellar
                                         systems: star clusters, galactic arms and rings"
\end{verbatim}

\vspace{1cm}

\begin{center}
{\Large \bf Effect of the Environment on the Fundamental Plane  \\ of
Elliptical Galaxies} \\

\vspace{0.5cm}

{\large \it E.A.~Evstigneeva, V.P.~Reshetnikov, N.Ya.~Sotnikova} \\
{\large \it St.Petersburg State University, Russia} \\
\end{center}

\vspace{0.5cm}

\rm
{\bf
We present an analysis of interacting E/S0 galaxies location on the
Fundamental Plane. Using the NEMO package, we performed N-body
simulations of close encounters and mergers between two spherical
galaxies. We followed how structural and dynamical parameters
(central density, half-mass radius and velocity dispersion) of
galaxies are changed during the encounter. We analysed the dependence
of these changes on initial mass concentration and presence of dark
halo. The results of our simulations are used to discuss the
Fundamental Plane for interacting early-type galaxies.}

\section{Fundamental Plane}

\rm

The Fundamental Plane (FP) defines one of the most important relationship
for early-type galaxies. The FP combines surface photometry parameters
($R_e$ -- effective radius and $\mu_e$ -- effective surface brightness
or $\langle \mu \rangle_e$ -- mean surface brightness within $R_e$)
with spectroscopy characteristics (line-of-sight central velocity
dispersion $\sigma_0$). The measured values of $R_e$, $\mu_e$ and
$\sigma_0$ for a sample of E and S0 galaxies do not fill this
3-parameter space entirely but rather a thin plane within it
(with scatter $\sim$ 0.1 dex). The FP can be projected onto any two
axes. Examples of these projections are the Kormendy relationship
($\mu_e$--log$R_e$), and the Faber-Jackson relationship between
luminosity and velocity dispersion.

The interpretation of the FP is still a matter of discussion.
The most popular point of view is that the FP is simply a
consequence of the Virial theorem and the fact that E/S0 galaxies
have similar mass-to-luminosity ratios and homologous structure
at a given luminosity.

The FP have several important implications. For instance, it can
be used to study the evolution of galaxies as a function of redshift;
the FP is a powerful tool for deriving redshift independent
distance estimates; etc.

There are no {\it clear} evidences that either the slope, zero-point
or scatter of the FP are dependent on spatial environment (see
discussion in Pahre et al.,
AJ 116, 1606, 1998). The purpose of the present note is to
explore possible influence of {\it strong encounters} of early-type galaxies
on their general characteristics and on the location within the FP.

\section{Fundamental Plane for Interacting and Merging Galaxies}

In order to study the FP for intensively interacting early-type
galaxies we have considered several pieces of observational data.

Fig.1 presents the Faber-Jackson relation for E/S0 galaxies
belonging to 1) VV or Arp systems (28 galaxies, stars),
2) CPG or TURNER binary systems, and 3) Hickson compact groups
(circles) according to the Hypercat Database 
(Prugniel \& Maubon, astro-ph/9909482). 

\begin{figure}[!ht]
\centerline{\psfig{file=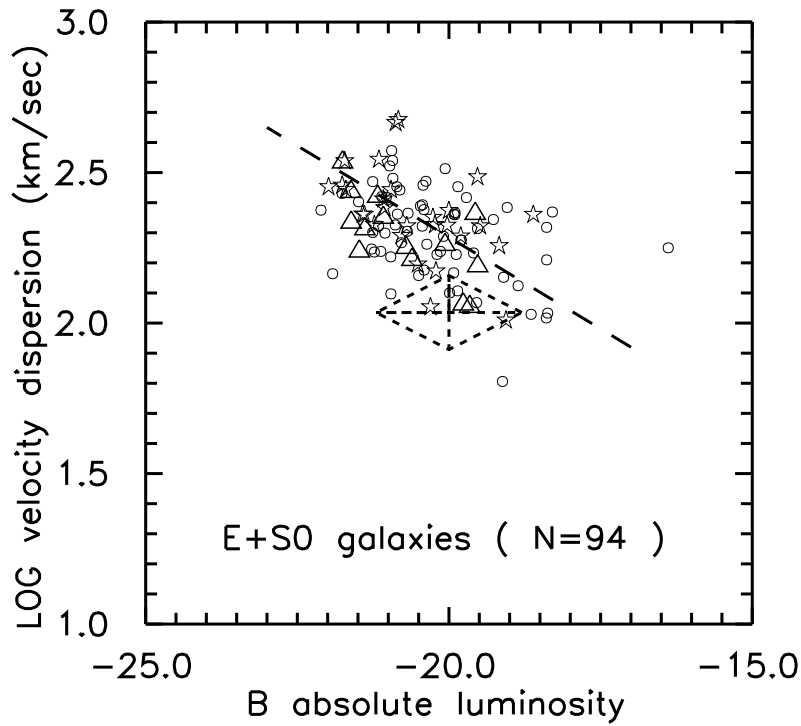,width=7cm}}
\caption{The Faber-Jackson relation for 94 interacting E/S0 galaxies.}
\end{figure}

Dashed line in Fig.1 shows the mean relation for 594 E/S0 galaxies
according to McElroy, ApJS 100, 105, 1995: $L_B \propto \sigma_0^{3.3}$.
The slope and the scatter ($\sigma({\rm lg}\sigma_0)=0.15$)
of the data for interacting galaxies are the same as for normal
E/S0.

Triangles in Fig.1 present the characteristics of 14 remnants of
disk-disk mergers (Keel \& Wu, AJ 110, 129, 1995).
As one can see, the mergers follow the same relation as early-type
galaxies but with small shift ($\sim 0.^m6$) to brighter absolute
luminosities. The dashed rhomb shows the mean value ($\pm 1 \sigma$)
for 9 starbursting infrared-luminous galaxies (Shier \& Fischer,
ApJ 497, 163, 1998). All these galaxies are in some stage of merging.
Starbursting mergers demonstrate significant brightening
($\sim 2^m$) in comparison with normal galaxies. One can expect
that relics of disk-disk mergers and infrared-luminous mergers
will evolve into normal ellipticals at the FP.

\begin{figure}
\centerline{\psfig{file=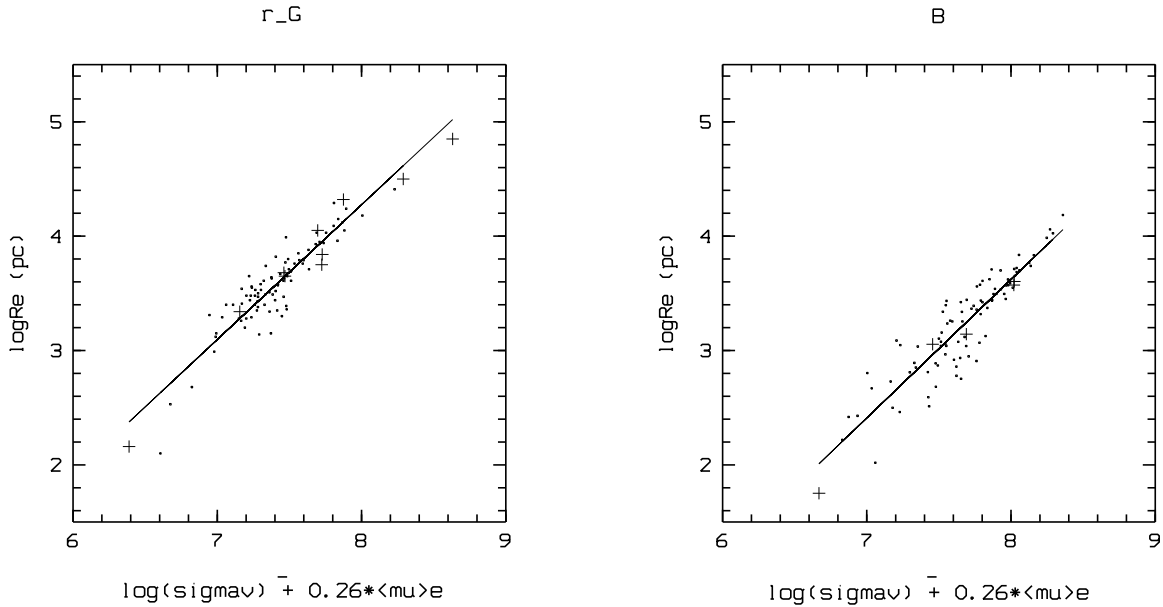,width=15cm,angle=-90}}
\caption{The FP for interacting E/S0 galaxies.}
\end{figure}

Fig.2 compares the FP location of ellipticals/lenticulars belonging to
VV and Arp systems (crosses) with relatively isolated galaxies
(points) (the data in the $r_G$ passband are form Djorgovski \& Davies,
ApJ 313, 59, 1987; in the $B$ filter -- Bender et al., ApJ 399, 462, 1992).
The FP scatter for interacting galaxies is the same (or even smaller) as for
non-interacting objects.

One can conclude that present observational data about the
global characteristics of strongly interacting E/S0 galaxies show
that the environment has no or almost negligible effect on the
FP. The only difference we see for forming (or young) ellipticals --
they are significantly brighter (for fixed $\sigma_0$)
due to superimposed burst of star formation.

\section{Numerical Modeling of Interacting Ellipticals}

In order to check our empirical findings we have performed
numerical modeling of close encounters of two ellipticals.

We simulated the evolution of the self-gravitating spherical galaxies by
using the NEMO package (Teuben, PASP Conf. Ser. 77, 398, 1995). This is a
freeware package designed to numerically solve gravitational N-body
problems. It consists of subroutines for specifying initial configurations
of stellar--dynamical systems (including many standard models) and
subroutines for simulating the evolution of these systems based on various
numerical schemes. In our computations, we used the scheme for constructing
an ''hierarchical tree'' (Barnes \& Hut, Nature 324, 446, 1986).

We considered two models for encounting galaxies. One of them is Plummer's
spherically-symmetric model

\begin{eqnarray}
\Phi(r) &=&
- \frac{G\, {\rm M}}{(r^2+a^2)^{1/2}} \, ,
\nonumber\\
&&
\nonumber\\
\rho(r) &=&
\frac{3 M}{4 \pi} \, \frac{a^2}{\left(r^2 + a^2\right)^{5/2}} \, ,
\nonumber
\end{eqnarray}
where M is the galaxy mass and $a$ is a scale length.

In some experiments galaxies are modeled by potential-density pair proposed
by Hernquist for spherical galaxies (ApJ 356, 359, 1990)

\begin{eqnarray}
\Phi(r) &=&
- \frac{G\, {\rm M}}{r + a} \, ,
\nonumber\\
&&
\nonumber\\
\rho(r) &=&
\frac{M}{2 \pi a^3} \, \frac{a^4}{r\,(r+a)^3}\, .
\nonumber
\end{eqnarray}

It is well-known that both models are described by distribution function
in analytical form and in the absence of numerical errors and dynamical
instabilities remain time-stationary.

The number of particles used in our numerical simulations ranged from
$N = 20\,000$ to $N = 50\,000$ per each galaxy. In this case, we managed to
suppress substantially the effects of pair relaxation and to trace the
evolution of merging galaxies on time scales up to
$t \approx 0.5\times10^9$ years.

We specified the initial distance between galaxies $R = 37.3$ kpc and chosed
the initial relative velocity in the range $V = 77.3-103.6$ km/s.
As a result we have a close encounter with merging (the distance of the
first closest approach was 5.2 kpc) and a distant encounter without merging
(in this case the minimum galaxy separation was 10.3 kpc). Fig.3 presents
some ``snapshots'' of a close encounter, showing the initial condition
($t=0$), the configuration near first maximum overlap ($t=30$), the
configuration shortly before the final merger ($t=34$), and a merger state
($t=40$).

\begin{figure}
\centerline{\psfig{file=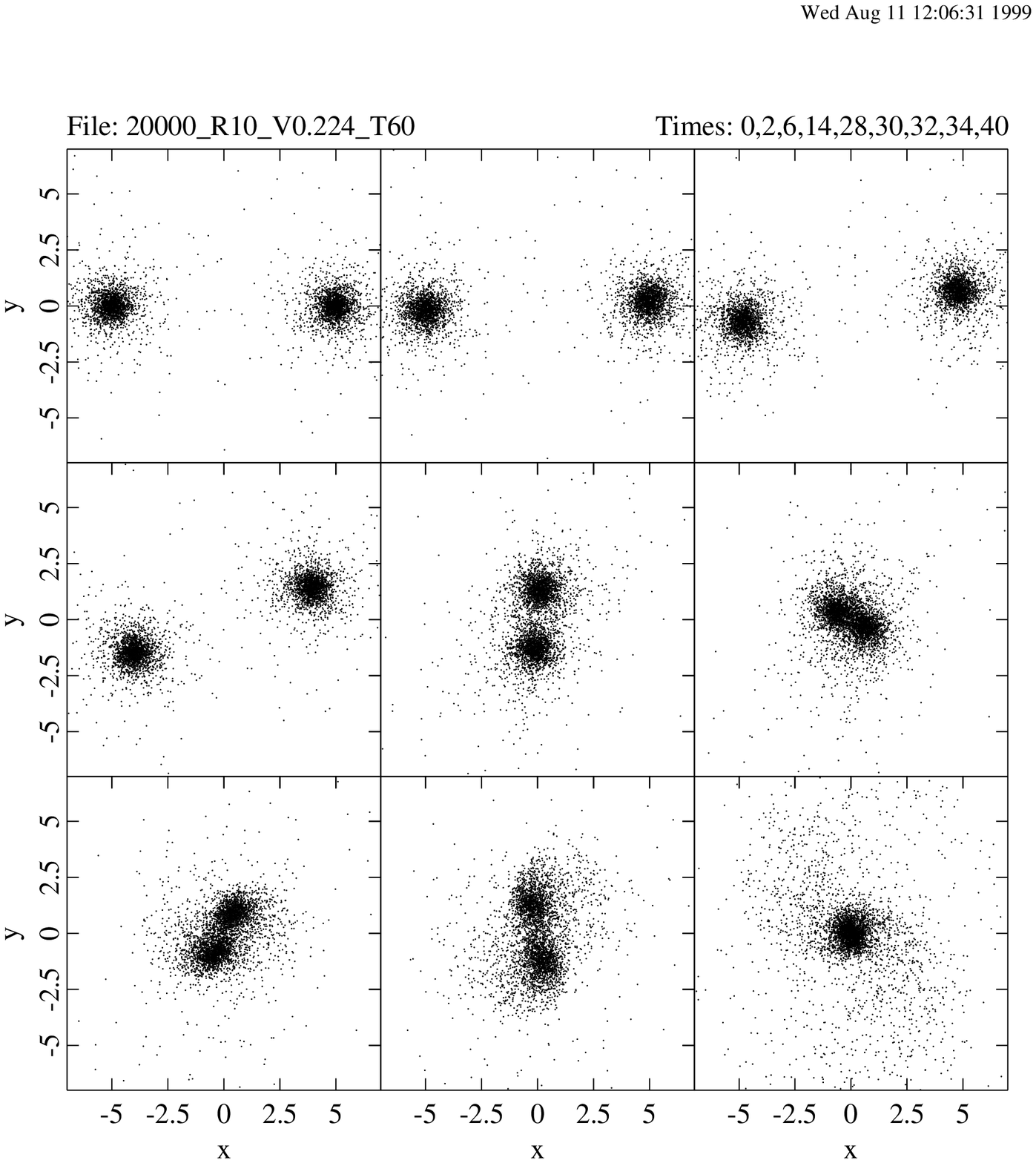,width=17cm}}
\caption{Close encounter of two identical Plummer's spheres.}
\end{figure}

Results are presented in the following system of units: gravitational
constant, $G = 1$, the galaxy mass, $M = 1$, the total energy of a sphere,
$E= -1/4$ (for Plummer's sphere $E= 3 \pi G M^2/ 64 a$, for Hernquist's
sphere $E= G M^2/ 12 a$). Scaled to physical value appropriate for a typical
elliptical galaxy, i.e. $M = 10^{11}~M_{\odot}$ and half-mass radius
$r_{1/2} = 3$ kpc ($r_{1/2}\approx1.31 a$ - for Plummer's sphere,
$r_{1/2}\approx2.41 a$ - for Hernquist's sphere), units of distance, time
and velocity are 3.73 kpc, 10.3 Myr, 345.3 km\,s$^{-1}$, respectively.

We followed the changes of central density, half-mass radius and central
velocity dispersion ($\sigma_0$). During the distant encounter all these
parameters were not altered. As to the close encounter, in this case the
most drastical changes in parameters were observed at the moment shortly
before the final merger. It appears the range of parameter changes depends
on initial mass concentration of a model (the Hernquist's sphere is more
concentrated than Plummer's model). Amplitudes of relative changes of 
the parameters for two models are:\\
Plummer's sphere --
$\delta r_{1/2}$ $\approx$ 67$\%$, $\delta \sigma_0$ $\approx$ 17$\%$, \\
Hernquist's sphere --
$\delta r_{1/2}$ $\approx$ 33$\%$, $\delta \sigma_0$ $\approx$ 6$\%$.

In some experiments we considered an encounter of two spherical galaxies
with dark matter components. The effect of parameter changes was less
pronounced than for the experiments without dark halo.

Our numerical experiments and observational data
show that global parameters of early-type galaxies are quite stable to 
even strong gravitational perturbation. Close encounters between galaxies 
does change the FP parameters ($R_e$, $\mu_e$, $\sigma_0$) 
within very limited time (a few $\times$ 10$^8$ years)
before the final merger directly. 

\end{document}